# Effects of Reducing Heat Treatment on the Structural and the Magnetic Properties of Mn:ZnO Ceramics

V. M. Almeida Lage,[a] R. T. da Silva,[a] A. Mesquita,[b]
M. P. F. de Godoy,[c] X. Gratens,[d] V. A. Chitta,[d] and H. B. de Carvalho[e]*

[a] *Universidade Federal de Ouro Preto – UFOP, 35400-000 Ouro Preto, MG, Brazil*

[b] *Departamento de Física, Instituto de Geociências e Ciências Exatas, Universidade Estadual Paulista – UNESP, 13500-900 Rio Claro, Brazil.*

[c] *Departamento de Física, Universidade Federal de São Carlos – UFSCar, 13565-905 São Carlos, Brazil*

[d] *Instituto de Física, Universidade de São Paulo - USP, 05508-090 São Paulo, Brazil*

[e] *Universidade Federal de Alfenas – UNIFAL, 37130-000 Alfenas, Brazil*

\* *Corresponding Author*: hugo.carvalho@unifal-mg.edu.br

**ABSTRACT**: Polycrystalline bulk Mn:ZnO ceramics with Mn nominal concentrations of 6, 11, 17 and 22 at.% were prepared trough solid-state reaction method and subjected to a heat treatment in reducing atmosphere (Ar (95%) and $H_2$ (5%)). The samples were studied with particular emphasis on their compositions, structural, and magnetic properties. A detailed microstructural and chemical analysis confirms the Mn doping of the wurtzite ZnO structure mainly at the surface of the ZnO grains. For the samples with higher Mn content, the secondary phases $ZnMn_2O_4$ and $Mn_{1-x}Zn_xO$ (Zn-doped MnO) were detected for the as prepared and the heat treated samples, respectively. The structural change of the secondary phases under heat treatment, from $ZnMn_2O_4$ to $Mn_{1-x}Zn_xO$, confirms the effectiveness of the heat treatment in to reduce the valence of the metallic ions and in the formation of oxygen vacancies into the system. In spite of the induced defects, the magnetic analysis present only a paramagnetic behavior with an antiferromagnetic coupling between the Mn ions. In the context of the bound magnetic polaron theory, it is concluded that oxygen vacancies are not the necessary defect to promote the desired ferromagnetic order at room temperature.

**KEYWORDS:** Multifunctionality. Spintronics. Zinc Oxide. Defect Engineering.

# 1 INTRODUCTION

ZnO is a true multifunctional material, it's a nontoxic abundant resource with superior environmental affinity, which is drawing much attention. With a wide bandgap (3.4 eV) and a strong binding energy (60 meV) at room temperature [1], the ZnO has been explored in different applications and technologies like in photocatalysis [2], as potential thermoelectric material in energy harvesting devices [3], in gas sensing systems [4] and also in photovoltaic elements [5], just to cite few. The functionalization of such materials is mainly achieved via defect engineering, by doping or even by the introduction of defects into the structural lattice in a proper and controlled way [6].

In such context, special attention has to be deserved to the magnetic properties of ZnO. It's a diamagnetic material and its magnetic functionalization takes place via doping with magnetic elements or by associating it to magnetic materials as in core-shell systems [7]. Magnetic ZnO at nanoscale has been explored in biomedicine as a bioimage agent or as a drug deliver [7], and also in antibacterial complexes [8]. Besides, magnetic ZnO is also a promising dilute magnetic semiconductor (DMS) to be used as a spin injection layer in spintronic devices. The research on transitional metal (TM) doped ZnO for spintronic purposes has been launched after the theoretical report by Dietl *et al*. [9]. According to these authors, wide band gap semiconductors, Mn-doped ZnO and GaN, would show Curie temperatures ($T_C$) above room temperature and presenting, therefore, a room temperature ferromagnetism (RTFM). However, in spite of the experimental and theoretical efforts of the last decades, the knowledge of about the magnetic properties of such materials, especially concerning the TM-doped oxides, are still controversial and inconclusive. By now, there is a consensus that the structural defects play an important role to tune the functional properties of the oxides, particularly their magnetic properties [10-14]. Considering the case of the TM-doped ZnO, one can find on the literature several reports connecting an observed RTFM to different structural point defects like zinc vacancies ($V_{Zn}$) [15, 16], zinc interstitial ($Zn_i$) [13, 17] and oxygen vacancies ($V_O$) [18, 19]. Among these defects, $V_O$ are widely believed to have a major role in promoting the desired RTFM [19-23]. A very often used technique to promote $V_O$ into the wurtzite ZnO ($w$-ZnO) structure is the heat treatment in low-oxygen [24, 25], inert [26, 27] or in reducing atmospheres [27-29]. However, under heating, the vaporization of ZnO takes place predominantly via dissociation into gaseous Zn and $O_2$ [30], and, therefore, heat treatments ends also promoting defects at Zn sites, and not only leading to $V_O$. For instance, Meng *et al*. [31] showed that Al and Mg co-doped $w$-ZnO under heat treatment in forming gas (97% $N_2$ and 3% $H_2$) at 650 °C for 1 h can completely remove the Zn from the samples, leaving behind a dominant Mg oxide phase.

From the theoretical point of view, different models have been proposed to explain the usual observed RTFM in DMS's. Here, the main accepted model for insulating systems is the bound magnetic polaron (BMP) theory proposed by Coey *et al*. [32], who argued that the ferromagnetic exchange among the TM doping elements is mediated by shallow donor electrons that form bound magnetic polarons, which overlap to create a spin-split impurity band. Nevertheless, high stable polarons can also be formed by holes bounded to acceptor defects into oxides [33, 34], and magnetic polarons can also be formed into insulating doped DMS's [35] and, for extension, into TM-doped $w$-ZnO [36]. Since $V_O$ are theoretically predicted to be a deep donor defect in the $w$-ZnO [37], $V_O$ can be associated to observed RTFM under the scope of the BMP model, as it has been done. However, $V_{Zn}$ is a shallow acceptor defect [37] and has also to be considered.

In this scenario, one can address the raised controversy mainly to the lack of knowledge in the nature and the control of the densities of defects in the studied materials. Therefore, the aim of the present report is to give further contribution to the understanding of the Mn incorporation into the *w*-ZnO structure, and shine some light over the question about the nature of the necessary defects to promote and stabilize the desired RTFM in ZnO-based DMS's. We conducted a detailed study of the microstructure and the magnetic properties of Mn:ZnO bulk samples with Mn nominal concentrations of 6, 11, 17 and 22 at.% prepared by the standard solid-state reaction method. The samples were prepared under oxygen atmosphere and, after all, fraction of the samples was subjected to a heat treatment in a reducing atmosphere of Ar (95%) and $H_2$ (5%) in order to introduce defects in the samples in a very controlled way. We found that, in spite of having induced $V_O$ via the reducing heat treatment, the samples only present a paramagnetic behavior with antiferromagnetic interaction among the Mn ions.

## 2   EXPERIMENTAL

Polycrystalline Mn:ZnO bulk samples with Mn nominal atomic concentrations of $x_N$ = 0.06, 0.11, 0.17 and 0.22 (6, 11, 17 and 22 at.%) were prepared via standard solid-state reaction method. Properly stoichiometric amounts of ZnO (Alfa Aesar 99.999% purity) and MnO (Alfa Aesar 99.99% purity) powders were mixed and ball milled for 5 h using Zn spheres. The resulting mixtures were cold compacted at a pressure of 600 MPa in the form of pellets (green pellets). The green pellets were finally sintered in oxygen atmosphere at 1400 °C for 4 h. These samples were labeled as "as prepared" (*AP*) samples. Fraction of these samples were then heat treated in reducing atmosphere of Ar (95%) and $H_2$ (5%) (Ar + $H_2$) at 600 °C for 3 h.  These samples were labeled as "heat treated" (*HT*) samples. Figure 1 presents a flowchart of the procedure employed in the preparation and characterization of the samples.

The structural characterization and the effects of heat treatment of the Mn:ZnO bulk samples were investigated by powder X-ray diffraction (XRD) using Rigaku Ultima IV equipment employing Cu K$\alpha$ radiation (30 kV, 40 mA, $\lambda$ = 1.5418 Å) recorded in the range of 2$\theta$ = 15 – 120° with steps of 0.02° at 7 s/step. The determination of the lattice parameters and the occupation factor over the structure were evaluated by using the Rietveld method implemented via General Structure Analysis System (GSAS) with the graphical user interface EXPGUI [38, 39]. The microstructure was determined using Scanning electron microscope (SEM)-LV JEOL JSM 6510 with resolution of 3 nm at 30 kV. The effective Mn concentration ($x_E$) incorporated into the hexagonal *w*-ZnO structure was estimated by energy dispersive X-ray spectrometry (EDS), using an Oxford XMAX 50 detector. Raman scattering spectroscopy was employed to study the incorporation of Mn and the resulting lattice disorder in the *w*-ZnO host structure, as well as to analyze the formation of segregated secondary phases. Raman measurements were acquired by using an IHR-550 Horiba Jobin Yvon spectrometer equipped with a Synapse Charge Coupled Device (CCD) detector, an Olympus BX41 microscope, and by using a 10× and 100× objective lenses in the backscattering configuration. The spectra were collected in several different points for each sample in order to enhance the statistical analysis. The excitation was performed with a 532 nm wavelength laser, with laser power of 1 mW over the sample. X-ray absorption spectroscopy (XAS) analysis was employed to determine the oxidation state (XANES – X-ray Near-Edge Spectroscopy) and to assess the environment (EXAFS – Extended X-ray Absorption Fine Structure) of the Mn atoms in the Mn:ZnO samples. These measurements were performed at the Mn and Zn *K*-edge in transmission mode using a Si (111) channel-cut monochromator at the XAFS2 beamline of the Brazilian Synchrotron Light Laboratory (LNLS). Finally, the temperature-dependent (3-300 K) magnetic measurements were performed by using a Cryogenics Superconducting Quantum Interference Device magnetometer (SQUID) in DC mode in magnetic fields up to 6 T.

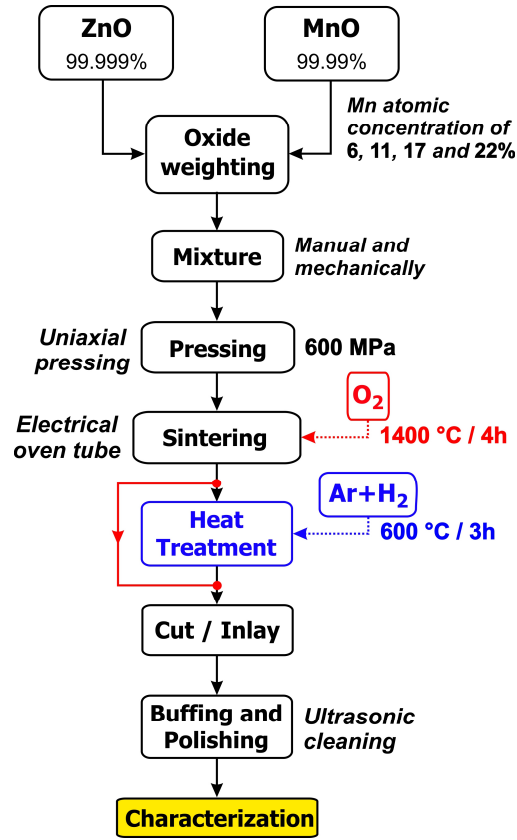

**Figure 1.** Flowchart of the preparation of the Mn:ZnO polycrystalline bulk samples.

## 3 RESULTS AND DISCUSSION

**3.1 *X-ray diffraction*:** Figure 2 shows the X-ray diffraction (XRD) results for the Mn:ZnO samples as a function of the Mn nominal content ($x_N$). The observed peaks correspond to those expected for polycrystalline hexagonal *w*-ZnO, space group $P6_3mc$ (PDF 01-072-8025) [40]. The relatively narrow line-widths revealed a highly crystalline quality for all samples. For the samples with $x_N$ = 0.06, *AP* and *HT*, it was not observed, within the detection limit of the measurements, any indication of additional phases. However, for the samples with $x_N$ greater than 6 at.% segregated phases were identified (symbols in Figure 2). For the *AP* samples the secondary phase zinc (II) dimanganate ($ZnMn_2O_4$) was identified (PDF 01-071-2499). After the heat treatment in the reducing atmosphere, samples *HT*, the $ZnMn_2O_4$ secondary phase almost disappears, and the manganese (II) oxide (MnO) secondary phase takes place (PDF 01-075-1090). The $ZnMn_2O_4$ phase has a tetragonal spinel structure ($AB_2O_4$ formula) in which the $Zn^{2+}$ occupies the $AO_4$ tetrahedral sites ($T_d$), and the $Mn^{3+}$ occupies the $BO_6$ octahedral sites ($O_h$). The crystal structure of $ZnMn_2O_4$ spinel belongs to the *I*$4_1$/*adm* space group. Besides, the MnO has a cubic rock-salt structure with $Mn^{2+}$ ions at octahedral sites, space group $Fm\bar{3}m$. This result indicates, as expected, a reduction of the oxidation state of the $Mn^{3+}$ to $Mn^{2+}$ with the heat treatment in the Ar + $H_2$ atmosphere.

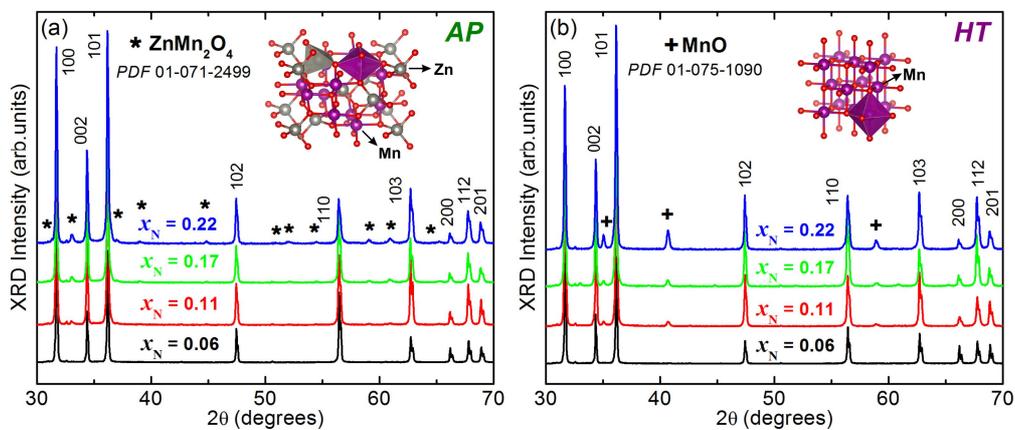

**Figure 2.** XRD pattern of the polycrystalline Mn:ZnO bulk samples prepared with different Mn concentration for the (a) *AP* and (b) *HT* samples. Stars show the peak positions of the segregated $ZnMn_2O_4$ phase in (a), and the crosses show the peak positions of the MnO segregated phase in (b).

The obtained Rietveld structural parameters are presented in the supplementary file (Figures S1, S2 and Tables S1 and S2). The main results concern the calculation of the fraction of the phases (*w*-ZnO, $ZnMn_2O_4$ and MnO) and the site occupation factors for each element in each phase. As the nominal concentration of Mn increases ($x_N$), the diffraction peaks corresponding to the secondary phases increase with respect to the peaks of the $w$-ZnO structure. The refined fraction of the phases revealed a linear behavior with the $x_N$ (Figure S3(a) in the supplementary file). We call attention to a small fraction of remain $ZnMn_2O_4$ phase into the *HT* samples (Table S2). The refined occupation factor for the Mn in the $w$-ZnO matrix, addressed as $x_R$, is presented in Table 1. It is stated that $x_R$ correspond to the Mn limit of solubility in the $w$-ZnO lattice under the preparation conditions used in this study. First, we observe that $x_R$ are in good agreement with the effective Mn concentration ($x_E$) measured via EDS (Table 1). We also observe that $x_R$ increases as the $x_N$ also increases. This result is corroborated by the linear increase of the refined volume of the $w$-ZnO unit cell as a function of $x_R$, as expected under the Vegard's law (Figure S3(b) in the supplementary file). The increase of the incorporated Mn concentration ($x_R$) into the $w$-ZnO will be further interpreted as a function of the changes in the grain growth conditions under the presence of higher amounts of the MnO precursor.

**3.2 *Electron Microscopy and Elemental Analyses*:** Figure 3(a) and Figure 3(b) show representative scanning electron microscope (SEM) images acquired over the polished surface of the Mn:ZnO samples. The images were acquired using a backscattered electron detector (BSE) in order to highlight the secondary phases, $ZnMn_2O_4$ (*AP*) and MnO (*HT*). It is observed in the microscale range the increase of the number and the size of the secondary phases (dark gray grains in Figure 3(a) and Figure 3(b)) as the $x_N$ increases, in accordance with the XRD and Rietveld results. Series of full scans over large areas at the samples *AP* and *HT* for $x_N = 0.06$ reveal the presence of a very small amount of secondary phases that was not detected in the XRD (red arrows in Figure 3). This finding can explain, in part, the slightly lower values obtained for $x_R$ and $x_E$ as compared to the $x_N$ (Table 1). This observation also highlights the limitation of the XRD technique in detect secondary phases even in the microscale range. For instance, in the context of the developing true DMS's it is hardly important to exclude the presence of any source of magnetic secondary phases that could leads to misinterpretations about the magnetic behavior. Therefore, for this purpose it is necessary to conjugate different experimental techniques and use highly sensitive tools in nanometric scale.

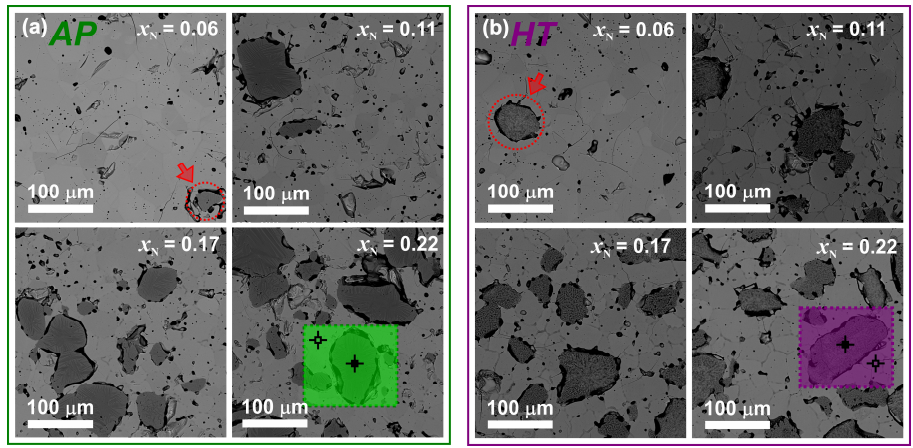

**Figure 3.** Representative scanning electron micrographs images obtained at backscattered configuration of the Mn:ZnO polycrystalline bulk (a) as prepared (*AP*) and (b) heat treated (*HT*) samples with the nominal concentrations of $x_N$ = 0.06, 0.11, 0.17 and 0.22. The secondary phase is well observed in the panels as dark gray grains. Highlighted areas in panels for sample $x_N$ = 0.22 correspond to the area were the EDS mapping were conducted (supplementary file, Figures S4 and S5).

Elemental analyses were performed via EDS measurements in multiple points and also over large surface areas of the samples. Representative EDS maps and point spectra measured at the Zn-rich (open cross in panel $x_N$ = 0.22 in Figure 3) and at Mn-rich (full cross in panel $x_N$ = 0.22 in Figure 3) grains are also presented in the supplementary file (Figures S4 and S5). The obtained average elemental percentages are listed in Table 1. We confirmed the Mn-doping of the *w*-ZnO lattice ($Zn_{1-x}Mn_xO$ in Table 1) for all samples. We observe that $x_E$ increases as $x_N$ increases and, as it was commented previously, $x_E$ is in good agreement with the $x_R$ factor obtained from the Rietveld analyses. An important result of this analyses concerns the chemical composition of the MnO secondary phase in the *HT* samples. Zn is also presented in the MnO phase (supplementary Figure S5(f)), what lead us to nominate this secondary phase more properly as $Mn_{1-x}Zn_xO$. As can be seen in Table 1, the Zn fraction into the MnO secondary phase matrix ($x_{Zn}$) is almost the same the Zn fraction into the $ZnMn_2O_4$ secondary phase in the *AP* samples (also labeled as $x_{Zn}$ in Table 1). This finding leads us to conclude that the performed heat treatment in reducing atmosphere (Ar + $H_2$) is effective in removing oxygen atoms from the samples. The $ZnMn_2O_4$ phase is, therefore, induced to a structural change to $Mn_{1-x}Zn_xO$ due to the high amount of oxygen vacancies ($V_O$) promoted under the heat treatment. In the same sense, but in different strength, we can infer that, with the heat treatment, $V_O$ were also promoted in the Mn-doped *w*-ZnO ($Zn_{1-x}Mn_xO$) main fraction of the samples.

**Table 1.** Mn concentration measured by EDS ($x_E$) and obtained via Rietveld refinement ($x_R$). The presented error for the $x_E$ data corresponds to the standard error of the mean. For the $ZnMn_2O_4$ phase in the as prepared samples (*AP*) and for the MnO phase in the heat treated samples (*HT*) we also presented the Zn concentration ($x_{Zn}$).

| Sample ($x_N$) | AP | | | | | HT | | |
|---|---|---|---|---|---|---|---|---|
| | $Zn_{1-x}Mn_xO$ | | $ZnMn_2O_4$ | | | $Zn_{1-x}Mn_xO$ | | $Mn_{1-x}Zn_xO$ |
| | $x_R$ | $x_E$ | Zn | Mn | $x_{Zn}$ | $x_R$ | $x_E$ | $x_{Zn}$ |
| 0.06 | 0.041(3) | 0.054(9) | 0.16(1) | 0.27(1) | 0.37(1) | 0.044(1) | 0.053(6) | 0.38(2) |
| 0.11 | 0.071(1) | 0.069(6) | 0.16(2) | 0.27(2) | 0.37(1) | 0.062(2) | 0.07(1) | 0.42(4) |
| 0.17 | 0.079(2) | 0.080(5) | 0.16(2) | 0.27(2) | 0.37(1) | 0.076(2) | 0.077(6) | 0.34(4) |
| 0.22 | 0.094(1) | 0.088(5) | 0.15(2) | 0.28(2) | 0.39(2) | 0.079(3) | 0.084(5) | 0.42(4) |

Table 2 presents the parameters obtained after a statistical analysis for the grain diameter distribution. The histogram for each sample is presented in the supplementary file (Figures S6 and S7). We observe that as $x_N$ increases the main diameter (*d*) of the grains decreases. It is well known that secondary phase particles can inhibit grain growth by pinning/dragging the migration of grain boundaries, which is often called as Zener effect [41]. Since it is observed a growing proportion of secondary phases ($ZnMn_2O_4$) in the *AP* samples with the increasing of the proportion of the MnO precursor, we can state that the formed $ZnMn_2O_4$ grains act as pinning/dragging center inhibiting the *w*-ZnO grain growth leading to a decrease of *d* as a function of $x_N$. With this result we can return to the observed increasing of the Mn effective concentration ($x_E$) into the *w*-ZnO matrix as function of $x_N$. We can infer here that the incorporation of the Mn ions into the *w*-ZnO takes place mainly at the grain surfaces. Increasing the MnO precursor proportion in the preparation of the Mn:ZnO ceramics, increasing $x_N$, increases the proportion of the $ZnMn_2O_4$ in the *AP* samples, leading to a decrease of the *w*-ZnO grain mean diameter (*d*), which, in turn, leads to an increase of the effective total surface area in the samples, allowing, therefore, a higher Mn incorporation. It is import to stress here that similar results were also reported for Co-doped nanograined *w*-ZnO samples [42].

**Table 2.** Particle size distribution analyses. *d* is the mean value of the particle diameter and $\sigma_g$ is the geometric standard deviation obtained by the log-normal fit of particle size distribution histograms for each sample. *N* is the total number of counted particles.

| Sample | AP | | | HT | | |
|---|---|---|---|---|---|---|
| ($x_N$) | *d* (μm) | $\sigma_g$ | N | *d* (μm) | $\sigma_g$ | N |
| 0.06 | 40.0(6) | 1.49(1) | 221 | 38.7(8) | 1.49(2) | 261 |
| 0.11 | 24.7(2) | 1.37(1) | 278 | 26.6(2) | 1.37(1) | 287 |
| 0.17 | 20.0(2) | 1.38(1) | 358 | 20.9(4) | 1.52(2) | 390 |
| 0.22 | 17.0(2) | 1.43(1) | 254 | 18.8(2) | 1.49(1) | 337 |

**3.3 *Raman scattering spectroscopy:*** To complement the structural analysis, Raman scattering analysis were also performed in order to conduct a detailed structural analysis and an evaluation of the chemical composition of the samples. Figure 4(a) shows a representative spectrum for the *AP* and *HT* samples with $x_N = 0.22$. The spectra for the other samples are quite similar to those shown in Figure 4(a). The Raman measurements were done here by using an objective of 10×, which leads to a relatively high analyzed average area. The presented spectrum corresponds also to an average of spectra acquired in several different points over the polished surface of the samples. The series of narrow modes centered at 97, 383 and 435 cm$^{-1}$ are assigned to the well-known $E_{2L}$, $A_1$(TO) and $E_{2H}$ *w*-ZnO modes, respectively [43]. In Figure 4(a) the spectra are normalized by the integrated intensity of the main mode $E_{2H}$. We also observe an extrinsic broad band between 500 and 600 cm$^{-1}$, which seems to be formed by the overlapping of several vibrational modes, with the most intense centered at 525 and 570 cm$^{-1}$. The mode at 570 cm$^{-1}$ can be attributed to the overlap of the $A_1(LO)$ and $E_1(LO)$ modes [44] for the $w$-ZnO lattice. These modes usually present low intensities in undoped $w$-ZnO systems due to the destructive interference between the deformation and the Frölich potentials [45]. However, the vibrational mode at 525 cm$^{-1}$ cannot be attributed to anyone of the $w$-ZnO structure. The emergence of this broad band is well report on the literature for undoped [46, 47] and doped $w$-ZnO with different elements [48-53], including the Mn [14, 54]. This band is clearly addressed to the structural disorder induced by the incorporation of dopants and defects into the $w$-ZnO lattice. Therefore, the observation of this extrinsic broad band for our samples is an indication that, at least, fraction of the Mn ions of the MnO precursor is incorporated into the $w$-ZnO lattice, corroborating the XRD, Rietveld and EDS analysis presented before. It is worth noting the higher relative intensity for *HT* samples. Considering that both samples have the same proportions of Mn (Table 1), we can state that the *HT* samples have a higher density of defects, such as $V_O$, than the *AP* samples, confirming the previous statement that the heat treatment in the reducing atmosphere (Ar + H$_2$) is effective in introducing those kinds of defects also in the Mn-doped $w$-ZnO lattice.

Figure 4(b) and Figure 4(c) shows representative Raman spectra of the segregated phases, ZnMn$_2$O$_4$ and Mn$_{1-x}$Zn$_x$O for the samples with $x_N$ = 0.22, respectively. The spectra were acquired with an objective of 100×, which allows a high spatial resolution. The obtained spectrum for the ZnMn$_2$O$_4$ phase at the *AP* samples is in quite good agreement with that reported for single phase ZnMn$_2$O$_4$ samples [55, 56]. At the Γ point of the Brillouin zone (BZ) the group theory predicts ten allowed optical phonons for the ZnMn2O4 represented by $\Gamma = 2A_{1g} + 3B_{1g} + B_{2g} + 4E_g$ [55]. The high-frequency modes (586, 631 and 679 cm$^{-1}$) are usually associated to the oxygen motion in the tetrahedral AO$_4$ sites, and the low-frequency modes to the octahedral BO$_6$ sites [57]. Nevertheless, further studies are necessary in order to corroborate this assumption. The Raman spectrum for the Mn$_{1-x}$Zn$_x$O phase at the *HT* samples (Figure 4(c)) quite resembles the spectrum for the ZnMn$_2$O$_4$ phase (Figure 4(b)). As presented before, MnO has a cubic rock-salt structure, and, therefore, it should not exhibit any first-order Raman activity. The observation of the almost the same spectrum as for the $ZnMn_2O_4$ reveals that the structural change from $ZnMn_2O_4$ to the $Mn_{1-x}Zn_xO$ under the heat treatment for some $ZnMn_2O_4$ grains is not complete. It means that there is some amount of $ZnMn_2O_4$, not detected via XRD, into, or close to, the $Mn_{1-x}Zn_xO$ grains in the *HT* samples. In fact, we could observe in some relatively large secondary phase grains in the *HT* samples a not complete phase change. The inset of Figure 4(c) shows a representative SEM image of secondary phase grain with a core-shell structure formed by a $ZnMn_2O_4$ core and a $Mn_{1-x}Zn_xO$ shell. To finalize, the main modes at around 322 and 679 $cm^{-1}$ of the $ZnMn_2O_4$ are also detected in the spectra presented in Figure 4(a). It is important to remember that in the region of 330 $cm^{-1}$ for the *w*-ZnO one can find also a relatively lower intensity *w*-ZnO vibrational mode $2E_{2L}$ at the M-point of the *BZ* [58].

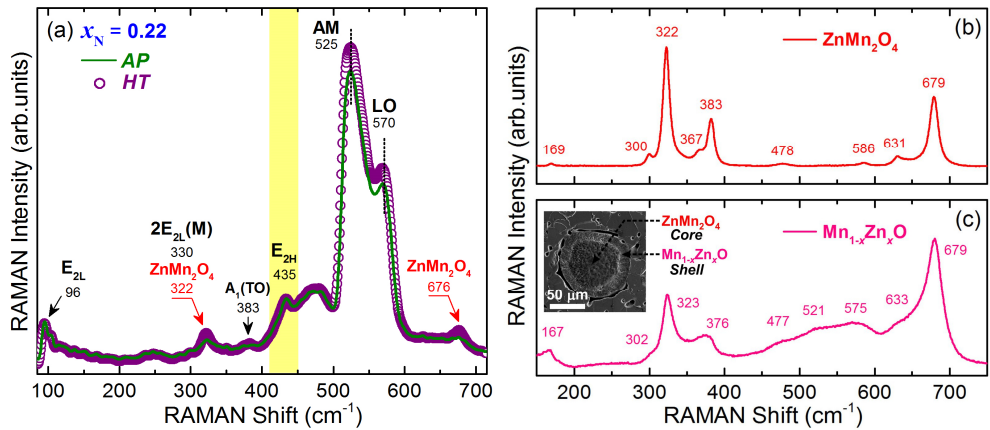

**Figure 4.** (a) Raman scattering spectra of the as prepared (*AP*) and heat treated (*HT*) Mn:ZnO polycrystalline bulk samples with the Mn nominal concentration of $x_N = 0.22$. The spectra were acquired at room temperature and are normalized by the main vibrational mode $E_{2H}$. The spectrum for the *AP* sample is shown in solid lines and the symbols correspond to spectrum for the *HT* sample. Raman scattering spectra of the sample with $x_N = 0.22$ (b) in the phase $ZnMn_2O_4$ (*AP*), and (c) in the phase $Mn_{1-x}Zn_xO$ (*HT*). The inset in (c) shows a representative SEM image of a relatively large core-shell grain formed by a $ZnMn_2O_4$ core and a $Mn_{1-x}Zn_xO$ shell. The phase identification was performed via elemental analysis (EDS).

**3.4 *X-Ray Absorption Spectroscopy:*** Figure 5 shows the XANES spectra obtained at room temperature for our Mn:ZnO bulk samples at the Mn absorption *K*-edge (6539 eV). It also presents the spectra for different reference Mn-based materials with different Mn oxidation states, metallic Mn, $MnO_2$ and the observed secondary phases $ZnMn_2O_3$ and MnO. A XANES spectrum gives information on the coordination symmetry and the valence of ions incorporated in a solid. The valence of the dopant ions can be analyzed by comparing their resulting edge structure to those obtained from reference samples. We observe for the samples *AP* and *HT* with $x_N = 0.06$ (Figure 5(a)) that the Mn ions assume mainly the oxidation state +2. The line shape of a absorption spectrum depends on the unfilled local density of states and the coordination symmetry of the absorbing element [12, 59]. Here we observe that the line shape for samples with $x_N = 0.06$ quite resembles previously reported spectrum for Mn-doped *w*-ZnO [14], indicating that the main fraction of the $Mn^{2+}$ ions occupy the $Zn^{2+}$ sites into the *w*-ZnO lattice (substitutional doping). It is also important to call attention to the relatively higher white-line (maximum after absorption edge) for the *HT* sample. This behavior is directly associated with an increase in the empty states left by open bonds due to vacancies defects promoted by heat treatment. This result corroborates the data obtained from Raman spectroscopy presented above concerning the introduction of defects via heat treatment in the Mn-doped *w*-ZnO lattice. By increasing $x_N$, it can be observed that the spectra for the *AP* samples approach the spectrum for the $ZnMn_2O_4$, not only in the edge, but also in the oscillations above the edge. In the same way, as $x_N$ increases, the spectra for the *HT* samples approach the spectrum presented for the reference MnO, as it was expected. Another relevant point concerns the observation of a small pre-edge absorption peak for all the samples (indicated at Figure 5(a)). This is only possible if $Mn^{2+}$ ions are located in sites without an inversion center of symmetry, like the tetrahedral $Zn^{2+}$ sites in the *w*-ZnO lattice [14, 51], instead of the octahedral sites in the $ZnMn_2O_4$ and MnO structures (structures in Figure 2). This result confirms the previous assumption that fraction of the Mn ions is taken the place of the $Zn^{2+}$ ions into the $w$-ZnO lattice even for the samples with $x_N > 0.06$.

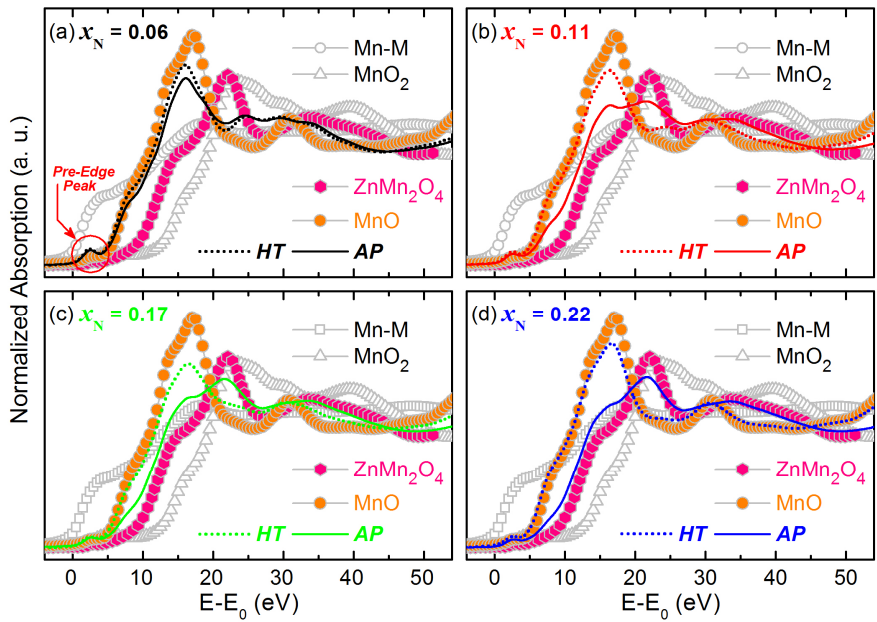

**Figure 5.** Mn $K$-edge XANES spectra for as prepared (*AP*) and heat treated (*HT*) Mn:ZnO polycrystalline bulk samples for the nominal Mn concentrations of (a) $x_N = 0.06$, (b) $x_N = 0.11$, (c) $x_N = 0.17$ and (d) $x_N = 0.22$. Spectra of metallic Mn (Mn-M), rocksalt MnO (valence +2), $ZnMn_2O_4$ (valence +3) and $MnO_2$ (valence +4) are also shown for comparison. $E_0 = 6539$ eV.

Figure 6 presents the extracted $k^3$-weighted Fourier transforms (FT) of our Mn:ZnO polycrystalline bulk samples, of the references powders $ZnMn_2O_4$ and MnO, as well as a spectrum obtained at Zn $K$-edge for an undoped ZnO reference sample prepared in the same condition as the *AP* samples. Qualitatively we observe that the spectra for the samples *AP* and *HT* with $x_N = 0.06$ are quite similar. By comparison we also see that the spectra of these samples are quite different from those obtained for $ZnMn_2O_4$ and MnO, otherwise, they correspond quite well to the spectrum acquired for the undoped ZnO reference sample at Zn $K$-edge. This result can let us infer again that the main fraction of the $Mn^{2+}$ ions in the *AP* and *HT* samples with $x_N = 0.06$ are located in the sites of the $Zn^{2+}$ ions in the $w$-ZnO lattice. For the samples with higher $x_N$ the case is different, exactly it was observed via XANES, for the *AP* samples, as $x_N$ increases, the obtained spectrum for each sample resembles more and more the spectrum for the reference powder ZnMn$_2$O$_4$. On the other side, for the *HT* samples, as $x_N$ increases, the spectra became similar to that for the reference powder MnO.

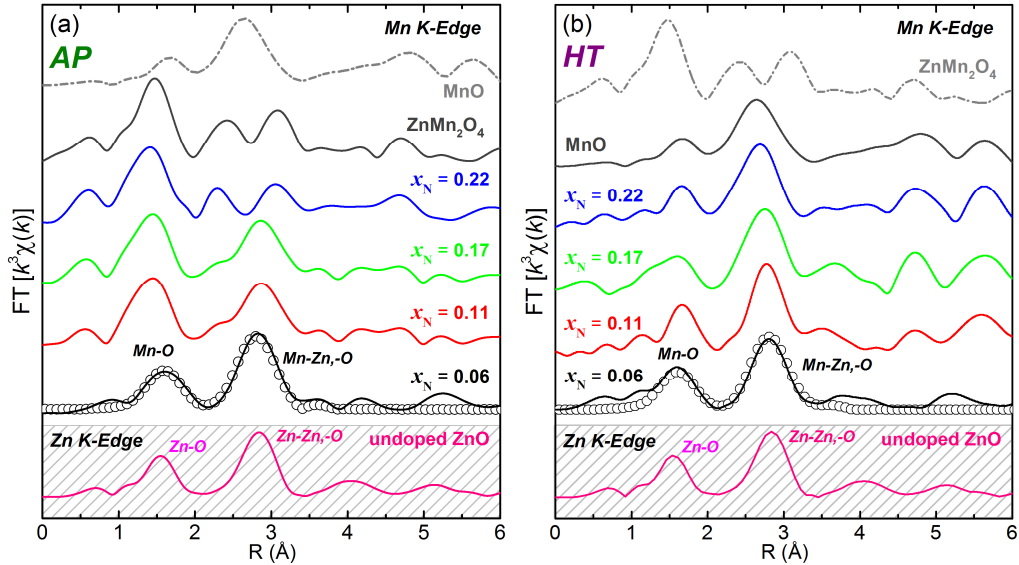

**Figure 6.** $k^3$ weighted Fourier transforms (FT) of Mn *K*-edge absorption data for the (a) as prepared (*AP*) and (b) heat treated (*HT*) polycrystalline Mn:ZnO samples and reference powders (ZnMn$_2$O$_4$ and MnO). The FT for the undoped ZnO sample were acquired ant Zn *K*-edge. The spectra are offset for clarity. Here the lines correspond to the experimental data and the symbols correspond to the theoretical simulation results.

A theoretical simulation of the measured FT for the *AP* and *HT* samples with $x_N = 0.06$ was also performed (Symbols in Figure 6). The technical details of the theoretical analysis were described in Ref. [14]. It was considered single and multi-scattering paths corresponding to the four successive atomic shells around Mn placed at the Zn sites of the ZnO structure according to the hexagonal wurtzite with P6$_3$*mc* space group. The parameters obtained via simulations are shown in Table 3. The obtained *QF* indicates the good reliability of the fits, which is confirmed by the comparison of the experimental (lines) and the fitted (symbols) data in Figure 6. These results also confirm that the Mn$^{2+}$ ions in the *AP* and *HT* samples with $x_N = 0.06$ are taking the place of the Zn$^{+2}$ ions in the *w*-ZnO lattice. An important and additional information can be extracted from the coordination number ($N$). For the first coordination shells $N$ decreases relatively from sample *AP* to *HT*, which indicates an increase of the vacancy defects as a function of the heat treatment, confirming also the previous assumptions about the efficiency of the heat treatment to induce these defects into the *w*-ZnO lattice.

**Table 3.** Mn $K$-edge EXAFS simulation results obtained by assuming $Mn^{2+}$ placed at $Zn^{2+}$ sites in the *w*-ZnO lattice for the samples *AP* and *HT* with $x_N = 0.06$. $R$ is the distance from the central atom, $N$ is the average coordination number, $\sigma^2$ the Debye-Waller factor and *QF* the quality factor.

| Sample ($x_N$) | Shell | $R$ (Å) | $N$ | $\sigma^2$ (Å$^2$) | *QF* |
|---|---|---|---|---|---|
| 0.06 (*AP*) | Mn-O | 2.04(1) | 4.3(6) | 0.007(2) | 3.02 |
| | Mn-Zn | 3.28(1) | 6.1(7) | 0.010(1) | |
| | Mn-Zn | 3.22(2) | 4.0(6) | 0.010(1) | |
| | Mn-O | 3.81(3) | 6(1) | 0.007(2) | |
| 0.06 (*HT*) | Mn-O | 2.03(1) | 3.8(6) | 0.005(2) | 4.01 |
| | Mn-Zn | 3.27(2) | 5(2) | 0.009(2) | |
| | Mn-Zn | 3.22(1) | 4(1) | 0.009(2) | |
| | Mn-O | 3.85(4) | 6(2) | 0.005(2) | |

**3.5** *Magnetometry:* The SQUID measurements were performed only for the *AP* and *HT* samples with nominal concentrations of $x_N = 0.06$ (free of secondary phases) and $x_N = 0.11$ (with low amount of secondary phases $ZnMn_2O_4$ (*AP*) and $Mn_{1-x}Zn_xO$ (*HT*)). All *M*(*H*) curves (not shown) present a present a paramagnetic (PM) behavior with a positive slope, rather characteristic of paramagnetic behavior (PM). No trace of ferromagnetic order is detected even at low temperatures. However, as we have undoubtedly shown in the previous sections, $V_O$ are induced in our samples via the heat treatment in the reducing atmosphere (Ar + $H_2$), and, in spite of that, only a PM behavior is observed. What lead us to conclude that, at least for the Mn-doped *w*-ZnO system, $V_O$ are not the necessary defects in order to achieve the RTFM. Following the results reported by Kittilstved *et al.* [60], Mn-doped *w*-ZnO presents RTFM only under *p*-type conduction, therefore, since $V_{Zn}$ is a shallow acceptor defect and can also form polarons, we can state that $V_{Zn}$ is a better candidate to promote the ferromagnetic coupling among the $Mn^{2+}$ ions.

The inverse of the *dc* magnetic susceptibilities ($\chi^{-1}$) as a function of temperature are presented in Figure 7. In the high temperature range (100-300 K), $\chi^{-1}$ presents a typical linear Curie-Weiss (CW) behavior. Below 100 K, $\chi^{-1}$ deviates from the linear dependence toward zero due to the effect of the ground states of Mn clusters coupled by antiferromagnetic interactions [61]. This feature was also reported for different TM-doped *w*-ZnO [62] and also for different DMS's [63-65]. In the Curie–Weiss law the *dc* magnetic susceptibility ($\chi$) as a function of the temperature (*T*) is given as $\chi(T) = C/(T-\theta)$; where *C* is the Curie constant per gram and $\theta$ is the Curie-Weiss temperature. The obtained parameters from the fitting of the experimental data in Figure 7 are presented in Table 4. A negative $\theta$ was obtained, which is a confirmation of a large antiferromagnetic (AF) interaction between Mn ions in the tested samples. We call attention to that only slightly differences can be observed between the *AP* and *HT* samples for both Mn nominal concentrations (Figure 7 and fit parameters in Table 4). This result indicates that the heat treatment does not affect considerably the magnetic properties of the samples, even considering the secondary phase transition from $ZnMn_2O_4$ (*AP*) to $Mn_{1-x}Zn_xO$ (*HT*) for sample with $x_N = 0.11$. In fact, both secondary phases present AF behavior among the Mn ions in their structures. Bulk MnO was one of the first studied AF materials using neutron scattering, and its magnetic properties can be easily found in textbooks with $\theta = -600$ K, and Néel temperature of $T_N = 118$ K [66]. Besides, the magnetic properties of the $ZnM_2O_4$ is not well completely understood, nevertheless, there is a consensus that it is also an AF material, with $\theta = -333$ K, and Néel temperature (still under debate), ranging from 15 K [57] up to 297 K [67]. Here, we could not observe in the susceptibility measurements for the sample *AP* with $x_N = 0.11$ any evidence of a PM-AF transition, most probably due to the low amount of the $ZnMn_2O_4$ secondary phases present in these samples (~3 %, supplementary material, Table S1). However, based on the available magnetic data for these materials, we can interpret the increase of the strength of θ for the *HT* samples, as compared to the *AP* samples, to the presence of MnO in these samples, which has a higher AF coupling among its Mn ions than the $ZnMn_2O_4$.

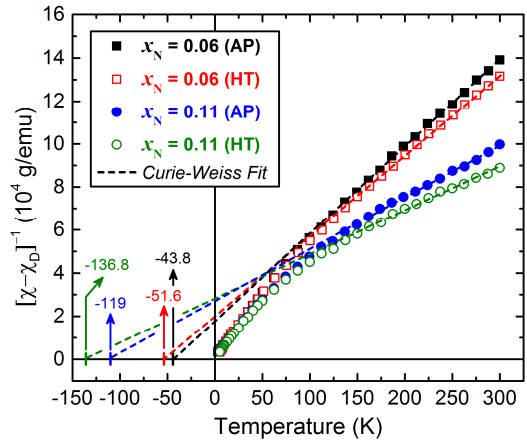

**Figure 7.** Inverse susceptibility as a function of the temperature for the as prepared (*AP*) and heat treated (*HT*) Mn:ZnO polycrystalline bulk samples with the Mn nominal concentration of $x_N$ = 0.06 and 0.11. The best fit of the high temperature data to Curie-Weiss law is shown as dashed lines. The diamagnetic background of the *w*-ZnO matrix ($\chi_D$) was subtracted from the raw data.

The Mn ion concentration ($x_M$), for the single phase samples ($x_N$ = 0.06), was determined from the *C* value obtained through the Curie Weiss fit and the relation $C = x_M N(g\mu_B)^2 S(S+1)/3k_B$, where *N* is the number of cation per gram, *g* and *S* are the Landé factor and the spin of the Mn ions (*g* = 2.0016, S = 5/2) [68], $\mu_B$ is the Bohr magneton and $k_B$ is the Boltzman constant. An estimation of the nearest neighbors exchange constant $J_1$ can be made by using: $\theta_0 = x_M 2zS(S+1)J_1/3k_B$, where *z* is the coordination number of the nearest neighbors (*z* = 12). The obtained values for $x_M$ are in good agreement with the measured Mn concentration presented before ($x_E$ and $x_R$). The $J_1/k_B$ values the are also consistent with previous results on $Zn_{1-x}Mn_xO$ bulk samples [69-71].

**Table 4.** Parameters determined from the magnetic measurements. $C$ is the Curie constant per gram and $\theta$ is the Curie-Weiss temperature, $x_M$ is the Mn concentration of the paramagnetic phase and $J_1/k_B$ is the effective exchange integral constant between the nearest-neighbors.

| Sample ($x_N$) | $C$ ($10^{-3}$ emu K g$^{-1}$) | $\theta$ (K) | $x_M$ (%) ($S = 5/2$) | $J_1/k_B$ (K) |
|---|---|---|---|---|
| 0.06 (AP) | 2,5 | −43.8 | 0.0456 | −13.7 |
| 0.06 (HT) | 2,2 | −51.6 | 0.0492 | −15.0 |
| 0.11 (AP) | 4.3 | −119 | - | - |
| 0.11 (HT) | 4.9 | −136.8 | - | - |

## 4 CONCLUSIONS

In summary, we have presented systematic microstructural and magnetic characterization of Mn:ZnO polycrystalline bulk samples prepared via standard solid-sate reaction method. Structural results confirm the incorporation of the Mn$^{2+}$ ions up to ~8.5 at.% (average value in Table 1) and indicate that it takes place mainly at the surface of the *w*-ZnO grains. For samples with higher Mn nominal content ($x_N$) the secondary phases ZnMn$_2$O$_4$ and Mn$_{1-x}$Zn$_x$O were detected for the *AP* and *HT* samples, respectively. It was directly confirmed that the heat treatment is efficient in reduce the metallic ions, Mn$^{3+}$ to Mn$^{2+}$, and in introduce $V_O$ into the system. For the sample of the lower $x_N$, in spite of the Mn$^{2+}$ doping of the *w*-ZnO lattice and the induced $V_O$ via heat treatment, the magnetometry results only show PM behavior with an AF coupling among the Mn$^{2+}$ ions, without any trace of the often reported RTFM. Once considering the BMP model, we can state that $V_O$ can not be addressed to the defect responsible for the magnetic coupling between Mn$^{2+}$ into the Mn-doped *w*-ZnO lattice.

**Acknowledgments:**

Support from agencies CNPq (grant 306715/2018-0) and FAPESP (07/56231-0; grant 2015/16191-5) is gratefully acknowledged. This research used resources of the Brazilian Synchrotron Light Laboratory (LNLS), an open national facility operated by the Brazilian Centre for Research in Energy and Materials (CNPEM) for the Brazilian Ministry for Science, Technology, Innovations and Communications (MCTIC). The XAFS2 beamline staff is acknowledged for the assistance during the experiments.